\documentstyle[12pt]{article}

\begin{document}
{\sf\begin{center}\noindent {\Large\bf Slow dynamo modes in compact Riemannian plasma devices from Brazilian spherical tokamak data}\\[3mm]

by \\[0.3cm]

{\sl L.C. Garcia de Andrade}\\

\vspace{0.4cm} Departamento de F\'{\i}sica
Te\'orica -- IF -- Universidade do Estado do Rio de Janeiro-UERJ\\[-3mm]
Rua S\~ao Francisco Xavier, 524\\[-3mm]
Cep 20550-003, Maracan\~a, Rio de Janeiro, RJ, Brasil\\[-3mm]
Electronic mail address: garcia@dft.if.uerj.br\\[-3mm]
\vspace{1cm} {\bf Abstract}
\end{center}
\paragraph*{}
Anti-dynamo modes are usually found in spheromaks plasma devices
experiments due to the fact that Cowling´s anti-dynamo theorem is
naturally applied to axisymmetric devices and flows. In this paper
full consideration is given to the existence of slow dynamo modes in
the case of compact Riemannian plasma devices without boundaries,
such as tokamaks, stellarators and torsatrons. It is shown that a
perturbed untwisted flow given by a decaying mode magnetic field is
able to generate a slow twisted dynamo plasma flow where the
unperturbed flow is also a steady flow. When the slow dynamo limit
is obtained for high Reynolds magnetic numbers $Rm$, the unperturbed
plasma flow achieve equilibrium as in spheromaks. A Riemann metric
of  a twisted thick magnetic flux tube with a very low aspect ratio
is used in the computations. The data of aspect ratio and other data
obtained from the Brazilian spherical tokamak at the National
institute for space research (INPE) are used to obtain a numerical
estimate for the maximum magnetic growth rate of the magnetic field
of the slow dynamo action as
${\gamma}_{1max}\approx{\frac{0.26}{a}{\times}\frac{v_{1{\theta}}}{B_{1s}}}$
for a toroidal initial magnetic field $B_{0}\approx{0.4T}$ and an
aspect data tokamak of $1.5$. For a tokamak internal radius of
$a\approx{0.2m}$ one obtains a maximum growth rate for the slow
dynamo as
${\gamma}_{1max}\approx{1.3{\times}\frac{v_{1{\theta}}}{B_{1s}}}$,
where $v_{1{\theta}}$ is the poloidal perturbed diffusive flow. The
safety factor is given by $q\approx{-1}$ which is $q<1$ implying
stability of spherical tokamak plasma. Slow dynamos have been
recently given in the literature as an example of Vishik's anti-fast
dynamo theorem [Phys Plasmas \textbf{15} (2008)], which can also
been tested in plasma experimental
devices.{\textbf{Key-words}:anti-dynamo theorems, slow dynamos.}

\newpage
 \section{Introduction}
 Recently Bellan \cite{1} has been discussed the existence of anti-dynamo modes in plasmas spheromaks.
 The stability has been given in terms of the safety factor conditions of instability $(q>1)$ or stability $(q<1)$. The Cowling's
 anti-dynamo theorem \cite{2}, which demands the existence of non-axisymmetric magnetic flows for the dynamo existence, can be circumvented in
 spheromaks as, even in axisymmetric spheromak devices, the existence of non-axisymmetric fluids and fields may exist, which per se, may guarantee
 dynamo action. Cowling \cite{2}, have show us how to recognise dynamo action in
 most types of magnetized plasma flows. The geometry of spheromaks, may be obtained
 by taking a very low aspect ratio, which is the ratio between the outer torus radius R and the cross-section a, or in the language of Riemannian
 twisted magnetic flux tube, the torus would be tick. Just as an example,in
tokamaks in general the aspect ratio can be built in plasma
laboratories, as
 $A=\frac{R}{a}={3}$, whereas in the Brazilian spherical tokamak \cite{3}, the ratio $A=1,5$. Actually magnetic plasma stability is favour for
 aspect ratios of $A<2$. Indeed, the magnetic topology of the field lines in the
 spherical tokamaks are highly
 torsioned at its center, which allows us these fields to reach values as high as $4T$. These torsion central lines are used to concentrate a strong
 magnetic field. This leads us naturally to use the Riemannian geometry of twisted magnetic flux tube \cite{4,5}
 to model these spherical tokamaks, provided one considers only thick flux tubes where the aspect ratio is low. Another advantage of its use is that
 this leads easily to plasma stability. Another advantage is that according to Schuessler \cite{6} astrophysical magnetic dynamos may be modelling by the
 flux tubes, as happens in the sun and other stars. Also another dynamo action mechanism called stretch-twist and fold (STF)
 as created by Vainshtein-Zeldovich
 \cite{7} may naturally appear in the magnetic lines of the spherical
 tokamaks, since a fundamental ingredient of twist is the Frenet
 torsion of its lines in the dynamo flow \cite{8}. From the
 mathematical side, using a dynamo model in Riemannian space has a
 long and solid tradition in toroidal maps as shown previously by
 Arnold et al \cite{9} and in more recent papers by the author
 \cite{5} to model plasma toroidal devices and astrophysical
 conformal dynamos. Another fundamental ingredient for dynamo existence, the folding, is a quantity which can be associated with
 Riemann curvature \cite{10} to provide the doubling of the magnetic
 field intensity by a repetition process that may guarantee dynamo action.
 In this paper untwisting flows are initially perturbed to yield slow dynamo action by using the spherical tokamak INPE data. In the presence of
 diffusion the relation between the growth rate of the perturbed magnetic field and the previous unperturbed one in terms of diffusion coefficient
 or resistivity. This allows us to show that a slow dynamo is obtained since the growth rate ${\gamma}$ vanishes when diffusion ${\eta}\rightarrow{0}$.
 Another Riemannian model for dynamo action has been recently obtained by Shukurov et al \cite{11} to model small-scale dynamos such as the Perm toroidal
 experimental dynamo \cite{12}. Another interesting distinction between the present model and the one addressed in reference (\ref{12}) is that in the case
 presently considered the perturbation of the magnetic field is not stationary. Recently a modification of Shukurov et al proposal was presented by the author can be considered as a thick Riemannian
 dynamo as well.
 Basic difference between our models and Moebius dynamo flow one proposed by Shukurov et al is that theirs is a numerical simulation and ours.
 Much earlier Mikhailovskii \cite{13} has used a non-diagonal Riemann metric to describe tokamaks and investigate the plasma instabilities. The paper is
 organized as follows: In section II the absence of diffusion is shown to lead to non-dynamos or marginal dynamos. In
 section III diffusion is turn on and the perturbation is shown to lead to a slow dynamo spherical tokamak. Discussions and future prospects are presented in section IV.
\newpage
\section{Riemannian plasma non-dynamo devices}
This section addresses the mathematical formalism of a general and
thick twisted Riemannian magnetic flux tube and show that a marginal
dynamo is obtained when a stationary unperturbed model is used when
the magnetic field is given along the magnetic lines, which is
$\textbf{B}=B(s)\textbf{t}$, and no slow dynamo is obtained. Here
one considers that the growth rate of magnetic perturbed and
unperturbed fields given respectively by $\textbf{B}_{1}$ and
$\textbf{B}_{0}$ are independently given by ${\gamma}_{1}$ and
${\gamma}_{0}$ where the magnetic fields are proportional to
\begin{equation}
|\textbf{B}_{0}|\approx{e^{{\gamma}_{0}t}} \label{1}
\end{equation}
and
\begin{equation}
|\textbf{B}_{1}|\approx{e^{{\gamma}_{1}t}} \label{2}
\end{equation}
where the perturbation process is given by
\begin{equation}
\textbf{B}=\textbf{B}_{0}+\textbf{B}_{1} \label{3}
\end{equation}
where $|\textbf{B}_{0}|>>|\textbf{B}_{1}|$. Here $\textbf{t}$ is
part of the Frenet vector frame along the curve coordinate given by
s-parameter. The complete Frenet frame is given by
$(\textbf{t},\textbf{n},\textbf{b})$ where vectors $\textbf{n}$ and
$\textbf{b}$ are the vectors that lay in the orthogonal plane to the
vector $\textbf{t}$ along the magnetic axis of the toroidal device.
In this section one shall consider the diffusionless case, and show
that this leads to a non-dynamo or at best to a marginal dynamo
where ${\gamma}_{0}$ vanishes. But before digging into the physics
of the problem let us take a moment to consider the Riemannian
geometry of flux tubes as given for the first time in the context of
solar plasma physics by Ricca \cite{4}. The magnetic flux tube
coordinates $(r,{\theta}_{R},s)$, is also used in plasma toroidal devices
called tokamaks. Since folding processes in
flux tubes can be represented by the Riemann curvature tensor,
destructive folding that leads to non-dynamos in diffusive media, can be obtained by
the vanishing of folding or vanishing of the Riemann curvature
tensor. General flux tube Riemannian metric is
\begin{equation}
d{s_{0}}^{2}=dr^{2}+r^{2}d{\theta}^{2}+K^{2}(r,s)ds^{2}\label{4}
\end{equation}
The thin Riemann-flat in twisted magnetic flux tube metric
is obtained by the constraining the relation $K^{2}:=(1-r{\kappa}(s)cos{\theta})$
to one. This is obtained as coordinate r approaches
zero. This happens in the neighbourhood of the torsioned flux
tube axis. Coordinate ${\theta}(s)$ is one of the Riemannian
curvilinear coordinates $(r,{\theta}_{R},s)$ and
${\theta}(s)={\theta}_{R}-\int{{\tau}(s)ds}$. The scalar function ${\tau}$ represents the
Frenet torsion. The thin tube metric is
\begin{equation}
{ds_{0}}^{2}=dr^{2}+r^{2}d{{\theta}_{R}}^{2}+ds^{2} \label{5}
\end{equation}
The torsion term is responsible for the
twist of the tube. Solar flux tubes are
closed in the inner parts of the Sun , and then the tubes can be considered as
compact Riemannian manifold without boundaries. Riemann gradient compact
operator is given in general diffusive substrate by
\begin{equation}
{\nabla}=\textbf{e}_{r}{\partial}_{r}+\frac{{\textbf{e}_{\theta}}}{r}{\partial}_{\theta}+\textbf{t}\frac{1}{K}{\partial}_{s}
\label{6}
\end{equation}
while general self-induction equation is
\begin{equation}
d_{t}\textbf{B}=(\textbf{B}.{\nabla})\textbf{v}+{\eta}{\Delta}\textbf{B}
\label{7}
\end{equation}
where in this section the resistivity ${\eta}=0$ in the ideal
plasma case. Throughout this paper, the magnetic field is strictly
confined along and inside the tube, which allows us to simplify the computations by considering that $B_{r}=0$ and that ${\partial}_{s}B_{s}=0$.
In the case consider here
\begin{equation}
{d}_{t}\textbf{B}={\gamma}\textbf{B}-{\tau}_{0}B_{0}\textbf{n}
\label{8}
\end{equation}
whose extra term is a non-inertial term similar to one that is
introduced into a inertial frame by the use of curvilinear
coordinates or Coriolis force in the frame. Therefore the diffusionless self-induction equation is given
by
\begin{equation}
[{\gamma}_{0}\textbf{t}-{{\tau}_{0}}^{2}\textbf{n}]B_{0}=B_{0}v_{0}{\partial}_{s}\textbf{n}
\label{9}
\end{equation}
where due to the highly torsioned character of the internal spherical tokamak, one has used the helical hypothesis of circular helices where the torsion
equals the Frenet curvature ${\kappa}_{0}$ and are constants. By comparison of the both sides of equation (\ref{9}) one obtains
\begin{equation}
{\gamma}_{0}=0
\label{10}
\end{equation}
and
\begin{equation}
v_{0}=-{\tau}_{0}
\label{11}
\end{equation}
Here the Frenet equations
\begin{equation}
d_{t}\textbf{t}={\kappa}\textbf{n}
\label{12}
\end{equation}
\begin{equation}
d_{t}\textbf{n}=-{\kappa}\textbf{t}+{\tau}\textbf{b}
\label{13}
\end{equation}
and
\begin{equation}
d_{t}\textbf{b}=-{\tau}\textbf{b}
\label{14}
\end{equation}
have been used to obtain the above results $v_{0}$ is the constant flow and the RHS of the equation (\ref{9}) represents the stretching of the flow. Actually $\textbf{v}=v_{0}\textbf{t}$
and the incompressibility condition of the flow
\begin{equation}
{\nabla}.\textbf{v}=0
\label{15}
\end{equation}
This implies actually that $v_{0}$ be constant. Thus from expression (\ref{10}) one must conclude that no dynamo action is possible in
for a constant modulus magnetic initial field in diffusionless media with a constant modulus stretching flow. Thus in the net section, we observe that by perturbing these
magnetic fields with a non-stationary unsteady magnetic field a dynamo action is present but the dynamo is still slow.
\section{Riemannian spherical tokamaks. slow dynamos and safety factor}
 The above perturbation scheme described in the last section, is
 used in the above magnetic self-induced equation, where
 $\textbf{B}_{0}$ and its perturbation $\textbf{B}_{1}$ obey the
 following zero and first-order equations
\begin{equation}
 d_{t}\textbf{B}_{0}=(\textbf{B}_{0}.{\nabla})\textbf{v}_{0}+{\eta}{\Delta}\textbf{B}_{0}
\label{16}
\end{equation}
and
\begin{equation}
d_{t}\textbf{B}_{1}=(\textbf{B}_{1}.{\nabla})\textbf{v}_{0}+(\textbf{B}_{0}.{\nabla})\textbf{v}_{1}+{\eta}{\Delta}\textbf{B}_{1}
\label{17}
\end{equation}
where ${\Delta}={\nabla}^{2}$ is the Laplacian Riemannian operator
given by
\begin{equation}
{\Delta}=[{{\partial}^{2}}_{r}+\frac{1}{r}{\partial}_{r}+\frac{{{\tau}_{0}}^{-2}}{r^{2}}\frac{(1+cos{\theta})^{2}}{cos^{2}{\theta}}{{\partial}_{s}}^{2}]
\label{18}
\end{equation}
Applying the following perturbed magnetic field $\textbf{B}_{1}$
into the above Riemannian Laplacian operator in the form
\begin{equation}
\textbf{B}_{1}=B_{1{\theta}}\textbf{e}_{\theta}+B_{1s}\textbf{t}
\label{19}
\end{equation}
where the dynamo mode m of the spatial toroidal coordinate-s appears
as
\begin{equation}
B_{1s}=b_{s}e^{{\gamma}_{1}t+ims} \label{20}
\end{equation}
where $m\in{\textbf{Z}}$, where $\textbf{Z}$ is the field of
integers numbers. Here $b_{s}$ is a constant with magnetic field
dimensions. Note that the poloidal magnetic field perturbation
$B_{1{\theta}}$ is given by the same expression with the only
difference that the constant $b_{s}$ of the toroidal field is now
replaced by $b_{1{\theta}}$. Note that the above choice seemingly
hide the coordinate-${\theta}$, but actually this is present in the
$ms$ exponent since the poloidal angular coordinate inside the
tokamak, does depend on coordinate-s as ${\theta}={\theta}(s)$ as
contained above. The reason one is not using a radial mode in the
magnetic field exponent is that one shall be considering here only
compact spherical tokamak surfaces, where the radial dependence of
the magnetic fields does not exist and the internal cross-section
radius of the tokamak is a. Bellow one shall show that the slow
dynamo mode m vanishes. Let us now display the expressions for the
Laplacian of the initial field $\textbf{B}_{0}=B_{0}\textbf{t}$
which by the solenoidality property
\begin{equation}
{\nabla}.\textbf{B}_{0}=0 \label{21}
\end{equation}
implies that $B_{0}$ does not depend on coordinate-s, since however
the tangent vector $\textbf{t}=\textbf{t}(s)$, the vector field
$\textbf{B}_{0}$ is a non-uniform unsteady magnetic field. Due to
the absence of radial dependence in the magnetic field over the
tokamak Riemannian compact surface without boundaries the Laplacian
of $\textbf{B}_{0}$ simplifies to
\begin{equation}
{\Delta}\textbf{B}_{0}=[\frac{(1+\cos{\theta})^{2}}{a^{2}{\cos^{2}{\theta}}}B_{0}(-\textbf{t}+\textbf{n})]
\label{22}
\end{equation}
From this expression, after some algebra, the complete diffusive
self-induction equation results in the following growth rate
\begin{equation}
{\gamma}_{0}=-{\eta}\frac{(1+cos{\theta})^{2}}{(acos{\theta})^{2}}
\label{23}
\end{equation}
Note that at this time, though the unperturbed magnetic field growth
rate does not vanish the situation is even worse here, cause
${\gamma}_{0}$ is negative and this means that the initial toroidal
magnetic field decays as happens with some primordial magnetic
fields in the universe \cite{14}. The other equation yields the
value of torsion ${\tau}$ as
\begin{equation}
{\tau}(s)\approx{\frac{{\eta}s}{a^{2}}}\label{24}
\end{equation}
This result seems to be rather interesting since it shows that
folding and twisting of the magnetic axis depends on a
straightforward manner from the Let us now show that this does not
happens with the perturbed field $\textbf{B}_{1}$ growth rate
${\gamma}_{1}$. The solenoidal property of the perturbed field
is
\begin{equation}
{\nabla}.\textbf{B}_{1}=0 \label{25} \end{equation} Application of
the appropriate magnetic field into this equation yields
\begin{equation}
\frac{B_{1{\theta}}}{B_{1s}}=cos{\theta} \label{26}
\end{equation}
and
\begin{equation}
im(cos^{2}{\theta}-1)B_{1s}=0 \label{27}
\end{equation}
this las equation leads to the constraint $m=0$ for the dynamo
action mode if it exists at all. Substitution of this mode into the
Riemannian Laplacian expression for the perturbed field reduces it
to
\begin{equation}
{\eta}{\Delta}\textbf{B}_{1}=-\frac{{\eta}^{2}(1+cos{\theta})^{2}}{a^{4}cos^{2}{\theta}}[\frac{sin2{\theta}}{2}-s]
B_{\theta} \label{28}
\end{equation}
This yields the following growth rate as
\begin{equation}
{\gamma}_{1}=\frac{{\eta}s}{a^{2}}\frac{B_{0}}{B_{1s}}sin{\theta}-\frac{2{\eta}^{2}(1+\cos^{\theta})^{2}}{a^{4}\cos^{2}{\theta}}[\frac{sin2{\theta}}{2}-s]\label{29}
\end{equation}
Note that in the limit ${\eta}\rightarrow{0}$ the growth rate
${\gamma}_{1}\rightarrow{0}$ which characterizes the slow dynamo
model. Thus one may say that the mode $m=0$ represents a slow dynamo
mode. More general modes may be found which may represent a fast
dynamo action in the spherical tokamak. Let us now consider the data
involved in the INPE brazilian spherical tokamak experiment, in the
case of maximum value of the growth rate ${\gamma}_{1max}$, by
taking into account the case when the magnetic Reynolds number $Rm$
is high or when the dynamo flow is highly conductive. In this case
the diffusion ${\eta}$ is small but finite and in this case the term
of second order in ${\eta}$ may be dropped and finally the last
expression reduces to
\begin{equation}
{\gamma}_{1max}\approx{\frac{2}{3}\frac{{v}_{1{\theta}}}{a}\frac{B_{0}}{B_{1s}}}
\label{30}
\end{equation}
Note that by using the INPE spherical tokamak data $a=0.2m$ and
initial toroidal field $B_{0}=0.4T$, this growth rate of slow dynamo
is
\begin{equation}
{\gamma}_{1max}\approx{\frac{0.26}{a}\frac{v_{1{\theta}}}{B_{1s}}}\approx{{1.3}\frac{v_{1{\theta}}}{B_{1s}}}
\label{31}
\end{equation}
here the torsion has been computed as
\begin{equation}
{\tau}_{0}\approx{\frac{1}{a}}\approx{3.3m^{-1}}\label{32}
\end{equation}
Finally let us compute the safety factor q of the tokamak as
\begin{equation}
q=\frac{d{\Phi}}{d{\theta}}=-\frac{1}{R}\frac{ds}{{\tau}ds}
\label{33}
\end{equation}
where we have used the expression above relating ${\theta}$ and
torsion integral $\int{{\tau}(s)ds}$. Since Frenet curvature is
given by
\begin{equation}
{\kappa}_{0}=\frac{1}{R} \label{34}
\end{equation}
These two last expressions yields
\begin{equation}
q=-\frac{{\kappa}_{0}}{{\tau}_{0}}\label{35}
\end{equation}
which in the circular helix case is given by $q=-1<1$ and thus
guarantees the tokamak plasma stability.

\newpage

\section{Conclusions} Anti-dynamo modes in several plasma devices
have been known in the literature. In this paper it is shown that
slow dynamo modes can be obtained in non-turbulent flows diffusive
plasma media. It is also shown that the absence of diffusion
forbides the presence of a fast dynamo action in the case of
initially toroidal flows where initial flows are aligned with the
magnetic field. Data from the brazilian spherical tokamak operating
in INPE is given to estimate the value of the growth rate of
perturbed flows magnetic field. It is shown that this depend upon
directly of the inverse of the magnetic Reynolds number which
displays an explicitly slow dynamo behaviour. The Riemannian flux
tube model in the thick case lead us to transform a toroidal tokamak
into a spherical one, and the growth rate of the perturbed field is
also proportional to the perturbed flow. When the perturbed flow
vanishes the growth rate vanishes as well and no dynamo action is
found whatsoever. Several other modes may be investigated in
spherical tokamaks with the hope fast dynamo action may be found in
future experiments. For example a more complicated model to INPE
tokamak would be given by a tube in the center of the spherical
tokamak where the torsion of the magnetic field could be confined
around a Riemannian torus.
\section{Acknowledgements} I am deeply greateful to Andrew D. Gilbert
and Renzo Ricca for their extremely kind attention and discussions
on the subject of this paper. Thanks are also due to I thank
financial supports from Universidade do Estado do Rio de Janeiro
(UERJ) and CNPq (Brazilian Ministry of Science and Technology).

  \end{document}